\documentclass[sigconf]{acmart}

\copyrightyear{2025} 
\acmYear{2025} 
\setcopyright{cc}
\setcctype{by}
\acmConference[CIKM '25]{Proceedings of the 34th ACM International Conference on Information and Knowledge Management}{November 10--14, 2025}{Seoul, Republic of Korea}
\acmBooktitle{Proceedings of the 34th ACM International Conference on Information and Knowledge Management (CIKM '25), November 10--14, 2025, Seoul, Republic of Korea}\acmDOI{10.1145/3746252.3760974}
\acmISBN{979-8-4007-2040-6/2025/11}

\AtBeginDocument{%
  }

\usepackage[hypcap=false]{caption}
\usepackage{array}
\usepackage{algorithm}
\usepackage{algorithmicx}
\usepackage{algcompatible}
\usepackage{balance}
\usepackage{mathrsfs}
\usepackage{mdwmath}
\usepackage{algorithm}  
\usepackage{cuted}
\usepackage{mathbbol}
\usepackage{mdwtab}
\usepackage[perpage, symbol*]{footmisc}
\usepackage{eqparbox}
\makeatletter
\@namedef{ver@lineno.sty}{9999/12/31}
\@namedef{opt@lineno.sty}{}
\makeatother
\usepackage{minted}
\usepackage{algpseudocode,float}
\usepackage{lipsum}
\usepackage{url}
\usepackage{tikz}
\usetikzlibrary{arrows.meta}
\usepackage{verbatim}
\usepackage{booktabs}

\usepackage{caption}
\usepackage{makecell}
\usepackage{subcaption}
\usepackage{graphicx}
\usepackage{textcomp}
\usepackage{xcolor}
\usepackage{url}
\usepackage{amsmath,amsfonts}
\usepackage{array}
\usepackage{color,framed}
\usepackage{multirow}
\usepackage{ulem}  
\usepackage{setspace}
\usepackage{comment}
\usepackage{pifont}
\usepackage{caption}       
\usepackage{ragged2e}      
\usepackage{xcolor}        
\definecolor{codebg}{gray}{0.92} 

\definecolor{KP-Agent}{HTML}{1f77b4}
\definecolor{ctr}{HTML}{ff7f0e}
\definecolor{regression}{HTML}{2ca02c}
\definecolor{impression}{HTML}{d62728}
\definecolor{cvr}{HTML}{9467bd}

\begin{document}

\title{KP-Agent: Keyword Pruning in Sponsored Search Advertising via LLM-Powered Contextual Bandits}

\author{Hou-Wan Long}
\email{houwanlong@link.cuhk.edu.hk}
\affiliation{%
  \institution{The Chinese University of Hong Kong}
  \city{Shatin}
  \country{Hong Kong}
}

\author{Yicheng Song}
\email{ycsong@umn.edu}
\affiliation{%
  \institution{University of Minnesota}
  \city{Twin Cities}
  \state{Minnesota}
  \country{United States}
}

\author{Zidong Wang}
\email{wangzidong@dsdigitalgroup.com}
\affiliation{%
  \institution{DS Digital Technology Group}
  \city{Hangzhou}
  \country{China}
}

\author{Tianshu Sun}
\email{tianshusun@ckgsb.edu.cn}
\affiliation{%
  \institution{Cheung Kong Graduate School of Business}
  \city{Beijing}
  \country{China}
}

\renewcommand{\shortauthors}{Hou-Wan Long, Yicheng Song, Zidong Wang, and Tianshu Sun}

\begin{abstract}
Sponsored search advertising (SSA) requires advertisers to constantly adjust keyword strategies. While bid adjustment and keyword generation are well-studied, keyword pruning—refining keyword sets to enhance campaign performance—remains underexplored. This paper addresses critical inefficiencies in current practices as evidenced by a dataset containing 0.5 million SSA records from a pharmaceutical advertiser on search engine Meituan, China’s largest delivery platform. We propose KP-Agent, an LLM agentic system with domain tool set and a memory module. By modeling keyword pruning within a contextual bandit framework, KP-Agent generates code snippets to refine keyword sets through reinforcement learning. Experiments show KP-Agent improves cumulative profit by up to 49.28\% over baselines.

\end{abstract}


\keywords{Sponsored Search Advertising, Agentic System, Reinforcement Learning}

\maketitle

\section{Introduction}
Sponsored search advertising (SSA) has become a dominant model in online advertising, contributing 28.7\% of the \$259 billion U.S. internet advertising revenue in 2024 as reported by the Interactive Advertising Bureau \cite{IAB_PwC2025}. SSA operates via a dynamic auction mechanism where advertisers select keywords and bid competitively; when user queries match these keywords, auctions determine ad placement and ranking on search engine results pages (SERPs) \cite{yuan2013real}. \textcolor{black}{The SSA ecosystem involves advertisers and search engines as the primary interacting parties, with search users influencing outcomes through feedback signals such as clicks and conversions.} Keywords serve as the critical connector between these entities \cite{yang2019keyword,sun2024value,han2020commercializing,han2019connecting}. Given the dynamic nature of user behavior and market competition, advertisers must continuously monitor performance and adjust keyword sets and bids in real-time to optimize results \cite{chatterjee2021sponsored,yang2022time}.

The adjustment of bidding prices and keyword generation have been well-researched areas in SSA \cite{tunuguntla2023bidding,kim2021should,jang2022multiple,cai2025rtbagent,joshi2006keyword,abhishek2007keyword,zhou2019domain,liu2014automatic,scholz2019akegis}. In contrast, keyword pruning, which refers to the process of refining keyword set to enhance campaign performance \cite{yang2023keyword}, is an often overlooked practice in the field of SSA research yet common in industry. \textcolor{black}{In SSA, advertisers often expand their keyword sets to reach broader audiences. However, keywords within the same campaign compete for a limited advertising budget. When the keyword set becomes too large, budget is spread thin, reducing the amount allocated to high-value keywords. As a result, overall efficiency may drop. To maximize value from a fixed budget, advertisers should regularly prune low-value keywords so that resources can be concentrated on those with better returns.} 

Few studies address keyword pruning directly. One notable work \cite{nagpal2021keyword} refines keyword sets by analyzing user search queries for semantic relevance and intent. While effective, this method relies on access to user search queries, which are typically not available to advertisers, as it is proprietary information owned by search engine companies. In the broader SSA domain, \cite{cai2025rtbagent} introduced RTBAgent, which adjusts keyword bids through multi-memory retrieval and a two-step decision process. However, RTBAgent focuses on bid optimization, not keyword adjustment. \textcolor{black}{Moreover, existing keyword pruning methods still rely on static heuristics in practice and struggle to adapt to evolving campaign dynamics (see Sec.\ref{related}). In contrast, LLM-based agentic systems offer flexible reasoning and autonomous interaction with dynamic environments \cite{yang2024multi,song2024ensemble}.} Therefore, to address this gap, we propose KP-Agent as shown in Fig.\ref{KP_Agent}, an LLM-based agentic system equipped with a toolset that encodes SSA domain knowledge and a memory module that provides few-shot examples and reflective feedback from past decisions. Unlike prior methods, KP-Agent operates solely on advertiser-side data (e.g., KPIs) and is specifically designed for keyword pruning. The contributions of this paper are as follows:

\begin{figure}[ht]
\includegraphics[width=0.45\textwidth]{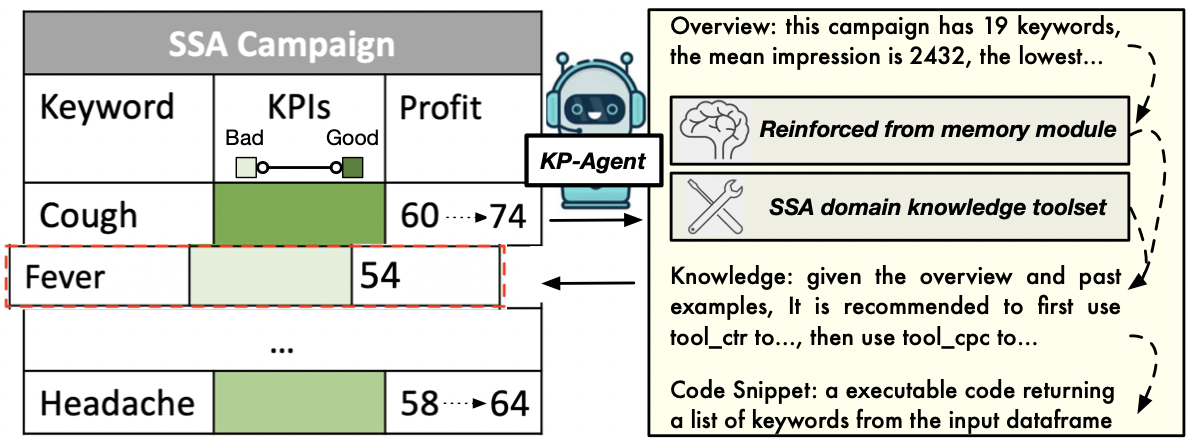}
        \caption{
        Keyword Pruning and KP-Agent.
        }
        \label{KP_Agent}
\end{figure}
\begin{itemize}
    \item[$\bullet$] We introduce KP-Agent, an innovative LLM agentic system that enhances the accuracy and efficiency of keyword pruning, addressing the current industrial challenges and academic gap.
    
    \item[$\bullet$] This study is pioneering in focusing on keyword pruning using advertiser-side data, eliminating the need for user search query data, which is typically inaccessible to advertisers as it is proprietary to search engine companies.

    \item[$\bullet$] We validate the effectiveness of KP-Agent through extensive experiments on Meituan's data. The results demonstrate that KP-Agent outperforms baseline methods in terms of the profit generated through keyword pruning.
\end{itemize}

\section{Motivation}
\label{related}
\begin{figure}[ht]
\includegraphics[width=0.45\textwidth]{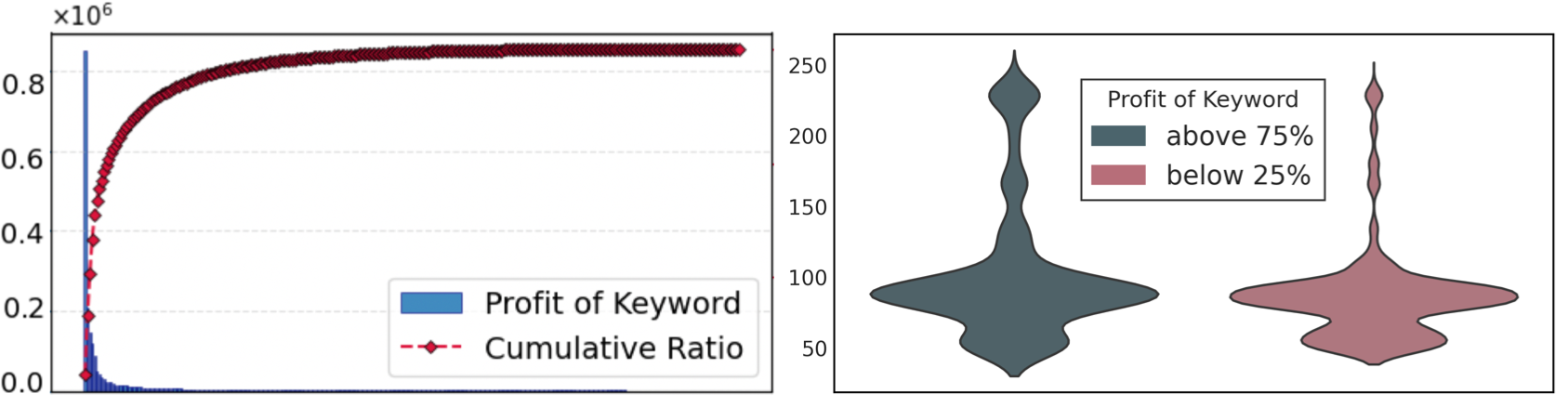}
\caption{
Empirical evidence from Meituan highlighting the need of effective keyword pruning method
}
\label{intro}
\end{figure}

We analyzed over 0.5 million SSA records from a pharmaceutical advertiser on Meituan, China’s largest on-demand delivery platform. The left plot in Fig.\ref{intro} shows that a small portion of keywords contributed to the majority of total profits, while the right plot shows similar bidding price distributions between top/bottom 25\% profitable keywords, which indicates low-value keywords crowded out high-value ones in budget allocation, underscoring the lack of an effective keyword pruning method. Moreover, among 45 campaigns over a 21-day period, only 29 keyword adjustments were made—just 4.6\% of the total. This infrequency of updates suggests advertisers failed to adapt to evolving performance signals. These findings underscore the urgent need for a more effective and adaptive keyword pruning solution such as KP-Agent.

\begin{table}[htb!]
  \caption{\textcolor{black}{SUMMARY OF NOTATIONS}}
  \centering
  \label{Notation}
  \scriptsize
  \begin{tabular}{c|l}
    \toprule
    Symbols & Descriptions \\
    \midrule
    $W_{i,t}$ & Keyword set for campaign $i$ on date $t$ \\
    $\lambda_{i,t}$ & KPIs for campaign $i$ on date $t$ \\
    $N_{\min}$ & Minimum required number of keywords per campaign \\
    $x_{i,t}$ & Context consisting of $W_{i,t}$ and $\lambda_{i,t}$ \\
    $a_{i,t}$ & Action selected by the agent (e.g., keyword pruning) \\
    $\mathcal{A}$ & Action space of keyword pruning operations \\
    $\pi$ & Policy mapping context to action \\
    $r_{i,t}$ & Reward obtained for campaign $i$ on date $t$ \\
    $p_{w,i}$ & Profit of keyword $w$ in campaign $i$ \\
    $o_{i,t}$ & Campaign overview derived from $W_{i,t}$ and $\lambda_{i,t}$ \\
    $\mathcal{O}(\cdot)$ & Template function generating $o_{i,t}$ \\
    $\mathcal{M}$ & Long-term memory module storing past examples \\
    $\mathcal{F}$ & Domain-specialized toolset for keyword pruning \\
    $\mathcal{E}_{i,t}$ & Retrieved $K$-shot examples from memory \\
    $\rho_{i,t}$ & SSA knowledge output from LLM after reasoning\\
    $C_{i,t}$ & Code snippet to be executed\\
    $\phi_{i,t}$ & Reflection generated from keyword pruning outcome \\
    $\text{LLM}_{\text{knowledge}}$ & LLM agent generating SSA-specific knowledge \\
    $\text{LLM}_{\text{code}}$ & LLM agent responsible for code generation and debugging \\
    $\text{LLM}_{\text{reflection}}$ & LLM agent producing market feedback reflection \\
    \bottomrule
  \end{tabular}
\end{table}

\section{Problem Formulation}
\label{formulation}

Building on the work of \cite{yang2023keyword}, we analyze an SSA system comprising \( M \) campaigns. For each campaign \( i \), we aim to select a subset of active keywords \( W_i' \subseteq W_i \) and corresponding budget allocations \( b_{w,i} \) for each \( w \in W_i' \), where the total cost does not exceed the campaign’s daily budget \( B_i \). Additionally, we impose an operational constraint that each campaign contains at least \( N_{\min} \) keywords to preserve exposure diversity.

\begin{equation}
\begin{aligned}
 \max  \sum_{i=1}^{M} \sum_{w \in W_{i}} p_{w,i} \text{ , subject to } |W_{i}| > N_{\min} &\text{ and } \sum_{w \in W_i'} b_{w,i} \leq B_i\quad\\
&\forall i \in \{1,2,\dots,M\}
\end{aligned}
\end{equation}

\section{KP-Agent's Solution Paradigm}
\label{method}
\textcolor{black}{We model keyword pruning as a series of independent decisions under uncertain market conditions using a Contextual Bandit framework} since the pruning decision depends primarily on the current campaign context. On each date $t$, the context \( x_{i,t} \) includes the keyword set \( W_{i,t} \) and KPIs \( \lambda_{i,t} \), such as impressions, click-through-rate (CTR), and conversion rate (CVR). KP-Agent selects an action \( a_{i,t} \in \mathcal{A} \) to prune \( W_{i,t} \) down to at least \( N_{\min} \), and receives an immediate reward \( r(x_{i,t}, a_{i,t}) \) reflecting profit and constraint satisfaction.

When KP-Agent encounters a keyword set \( W_{i,t} \) and its associated KPIs \( \lambda_{i,t} \), it uses its memory module \( M \) and tool set $\mathcal{F}$ to determine the action \( a_{i,t} \) through the subsequent process:
\begin{equation}
\label{policy}
\begin{tikzpicture}[baseline=(current bounding box.center)]
\node (eq) at (0,0) {
$\begin{aligned}
    o_{i,t} = \mathcal{O}(x_{i,t}) = \mathcal{O}(W_{i,t}, \texttt{Stats}(\lambda_{i,t}))& \\
    \rho_{i,t} = \text{LLM}_{knowledge}(o_{i,t}, \mathcal{E}_{i,t}, \mathcal{F})& \\
    C_{i,t} = \text{LLM}_{code} \left( \rho_{i,t} \right)& \\
    \pi_{\theta} (a_{i,t} | x_{i,t}, \rho_{i,t}) \equiv \text{Exe} \left( C_{i,t}, W_{i,t} \right)& \\
\end{aligned}$
};
\draw[->, thick] (3.2,1.2) -- (3.2,-0.8);
\end{tikzpicture}
\end{equation}

The agent's objective is to learn a policy \(\pi_{\theta}^{*}\) that maximizes the expected reward based on the current context \(x_{i,t}\). Action selection is guided by agent's internal reasoning output \(\rho\). The policy is optimized as follows:
\begin{equation}
\begin{aligned}
\pi_{\theta}^{*} = \arg \max_{\pi_{\theta}} \mathbb{E}_{\pi_{\theta}} \left[ \sum_{i=1}^M r(x_{i,t}, a_{i,t}) \mid x_{i,t}=x, \rho_{i,t}=\rho \right]\\
\text{s.t. } \pi(a_{i,t} \mid x_{i,t}, \rho_{i,t}) \text{ satisfy Eq.\ref{policy} } \forall i
\end{aligned}
\end{equation}

Algorithm~\ref{alg:ks_agent} illustrates the workflow of KP-Agent. The process begins with the analysis of each campaign's KPIs \(\lambda_{i,t}\), using the $\texttt{Stats}(\cdot)$ function to compute summary statistics. These statistics, along with the keyword set \(W_{i,t}\), are passed to the prompt template \(O(\cdot)\) to generate a structured overview of the campaign data, denoted as \(o_{i,t}\). Next, KP-Agent retrieves relevant few-shot examples \(\mathcal{E}_{i,t}\) from the memory module \(\mathcal{M}\) (see Sec.\ref{Memory}), and, together with the overview \(o_{i,t}\) and the information of toolset \(\mathcal{F}\) (see Sec.\ref{Toolset}), provides them to the \(\text{LLM}_{\text{knowledge}}\) agent. This agent applies domain-specific SSA knowledge to produce a reasoning output \(\rho_{i,t}\), named with knowledge, which guides the selection and application of tools suitable for the input context. 

\textcolor{black}{Since the input data is tabular, we avoid direct LLM inference to prevent hallucination \cite{dong2024large}. Instead, the knowledge \(\rho_{i,t}\) is passed to the $\text{LLM}_{code}$ agent, which generates a code snippet \(C_{i,t}\) that transforms the input table into a pruned keyword set as shown in Fig.\ref{toolset_code}}. This code snippet is executed via the $\texttt{CodeExecutor}$. If execution errors occur, the agent automatically debugs the code by refining it until successful execution is achieved. The resulting keyword set \(W'_{i,t}\) is used to perform an environment-compatible action. Finally, the \(\text{LLM}_{\text{reflection}}\) agent generates a reflection \(\phi_{i,t}\) based on the action taken and the resulting market feedback. This reflection, along with the campaign's overview, knowledge, and code snippet, is stored in the memory module \(\mathcal{M}\) for future retrieval. The toolset and memory module are described in detail in subsequent sections.

The reward for KP-Agent is derived from self-reflection on action and profit, ensuring self-alignment \cite{yuan2025selfrewardinglanguagemodels}. This ensures that KP-Agent's actions are closely aligned with the policy that maximizes the expected return on keyword pruning, considering the current state and decision-making insights at each step.

\begin{algorithm}
\scriptsize
\caption{Workflow of KP-Agent for Keyword Pruning}
\label{alg:ks_agent}
\begin{algorithmic}[1]
\State \textbf{Initialize:} 
    \Statex $\quad$\textbullet\ Tool set $\mathcal{F} \gets \{f_1,\dots,f_n\}$ 
    \Statex $\quad$\textbullet\ Memory module $\mathcal{M} \gets \mathcal{M}_0$
    \Statex $\quad$\textbullet\ Constraints $N_{\min}$ (min retention)

\For{$t \gets 1$ \textbf{to} $T$}
    \For{each campaign $i \in \{1, \ldots, M\}$}
        \State Observe state $s_{i,t} = (W_{i,t}, \lambda_{i,t})$ 
        \State Generate campaign overview:  $\quad o_{i,t} \gets \mathcal{O}(W_{i,t}, \texttt{Stats}(\lambda_{i,t}))$
        \State Retrieve similar examples: $\quad \mathcal{E}_{i,t} \gets \arg \text{TopK}_{\max}(\sin( o_{i,t}, o )| o \in \mathcal{M}))$
        \State Generate domain knowledge: $\quad \rho_{i,t} \gets \text{LLM}_{\text{knowledge}}(o_{i,t}, \mathcal{E}_{i,t}, \mathcal{F})$
        \State Generate code snippet: $\quad C_{i,t} \gets \text{LLM}_{\text{code}}(\rho_{i,t})$
        
        \Repeat
            \State Execute $C_{i,t}$ via \texttt{CodeExecutor}, get $W'_{i,t}$
            \If{execution error occurs}
                \State Parse error $e_{i,t}$, debug via $\text{LLM}_{\text{code}}$
                \State Update code snippet $C_{i,t}$
            \EndIf
        \Until{execution succeeds}
        
        \State Deploy $W_{i,t+1} \gets W'_{i,t}$
        \State Observe reward $r_{i,t} = \sum_{w\in W_{i,t+1}} p_{w,i}$
        \State Generate reflection: $\quad \phi_{i,t} \gets \text{LLM}_{\text{reflection}}(o_{i,t}, C_{i,t}, r_{i,t})$
        \State Update memory $\mathcal{M} \gets \mathcal{M} \cup \{(o_{i,t}, \rho_{i,t}, C_{i,t}, \phi_{i,t})\}$
    \EndFor
\EndFor
\end{algorithmic}
\end{algorithm}

\subsection{SSA Domain-Specialized Tools for Hallucination-Free Tabular Processing}
\label{Toolset}
\begin{figure}[ht]
\includegraphics[width=0.47\textwidth]{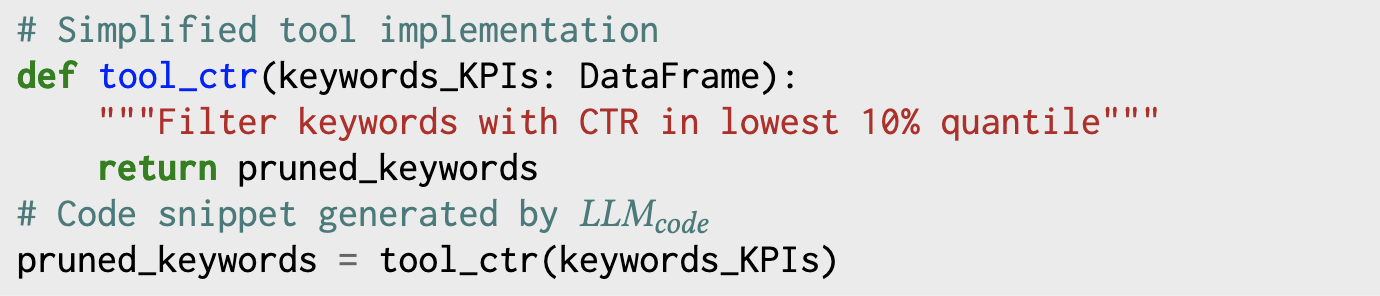}
\caption{
Simplified Demonstration of Tool and Code Snippet
}
\label{toolset_code}
\end{figure}

Each campaign's input data is structured in a tabular format, including columns such as date, keyword, and KPIs. Typically, to process such tabular data, LLMs convert dataframes into markdown tables and then input them as text \cite{fang2024largelanguagemodelsllmstabular}. However, since LLMs are primarily trained on textual data, they inherently struggle with accurately reasoning over tabular data, which can lead to "hallucinations" \cite{dong2024large}. To address this challenge, inspired by \cite{shi2024ehragent} and through collaboration with SSA industry experts, we developed a specialized toolset $\mathcal{F}$ as shown in Fig.\ref{toolset_code}. \textcolor{black}{Each tool is a function that performs a specific operation on the tabular, such as sorting and filtering.} This toolset encapsulates the principles of evaluating keywords under various settings, as determined by the data overview defined in Section \ref{method}. By converting these principles into functions, we provide a toolset for subsequent code generation. This approach allows key insights to be transformed into function outputs, thereby avoiding the pitfalls of direct tabular data reasoning by LLMs.

\subsection{Memory-Augmented Reflection for Few-Shot Learning}
\label{Memory}
\textcolor{black}{To enable few-shot learning in complex SSA scenarios, we implement a long-term memory module, denoted as $\mathcal{M}$, which stores past successful examples for reuse. Each entry in $\mathcal{M}$ includes a campaign's overview, knowledge, corresponding code snippets, and reflections on market feedback generated by $\text{LLM}_\text{reflection}$. KP-Agent retrieves the most relevant few-shot examples based on the similarity between the current campaign overview and those stored in memory. Overview similarity serves as a compact representation of campaign status, allowing KP-Agent to identify structurally similar past situations for effective knowledge transfer.} This mechanism supports dynamic adaptation akin to policy selection in a contextual bandit. The selection of \( K \)-shot examples \(\mathcal{E}(s_{i,t})\) is defined as follows: 
\begin{equation}
\mathcal{E}_{i,t} = \arg \text{TopK}_{\max}(sin(o_{i,t}, o )| o \in \mathcal{M}))
\end{equation}
where \(\arg \text{TopK}_{\max}(\cdot)\) identifies the indices of the top $K$ elements with the highest values from $\mathcal{M}$, while $\sin(\cdot, \cdot)$ calculates the similarity between two overviews using negative Levenshtein distance.

The retrieved examples are used to construct the prompt for $\text{LLM}_{\text{knowledge}}$, guiding code generation. Notably, each retrieved case includes a reflection based on previous outcomes, allowing the KP-Agent to refine its decision-making using accumulated SSA domain experience.

\begin{figure*}[!htb]
    \includegraphics[width=\textwidth]{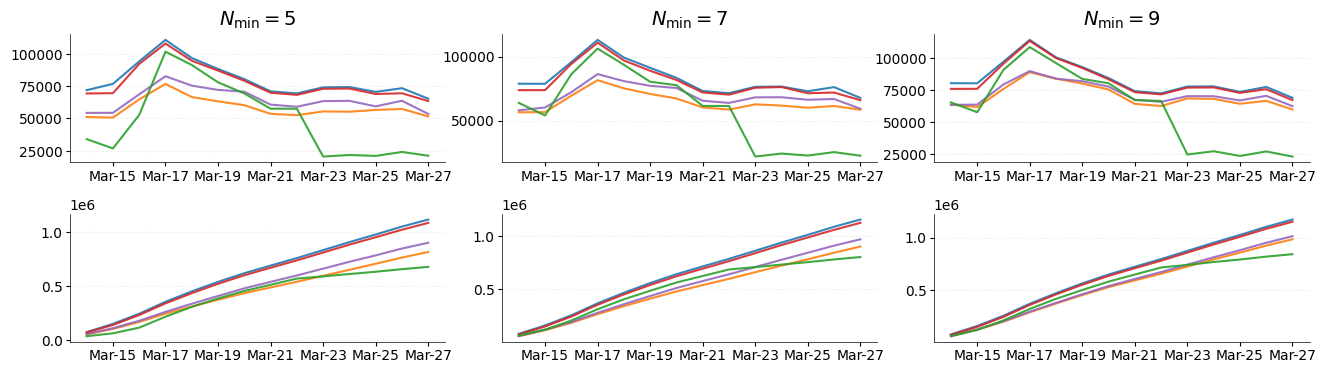}
    \caption{Performance comparison between KP-Agent and baseline methods exhibiting daily profit (first row) and cumulative profit (second row) across varying \(N_{\text{min}}\) values, where the different color represent the corresponding methods: \textcolor{KP-Agent}{$\blacksquare$} KP-Agent, \textcolor{impression}{$\blacksquare$} Impression-Rank, \textcolor{ctr}{$\blacksquare$} CTR-Rank, \textcolor{cvr}{$\blacksquare$} CVR-Rank, \textcolor{regression}{$\blacksquare$} Impression Regression.}
    \label{exp1}
\end{figure*}

\section{Experiment}
\subsection{Dataset}
Given the lack of suitable public benchmarks for evaluating keyword pruning, we've collected a real dataset from Meituan for our study. This dataset comprises real-time SSA data over 21 days in 2025. After filtering out campaigns with few keywords used for special promotions, it covers 45 advertising campaigns, 278 unique keywords, featuring 2.2 million impressions, 175,816 clicks, and a total advertising cost of 83,531 RMB. This data illustrates the market environment and offers a comprehensive view of user responses from the advertiser's perspective. Each dataset record comprises three elements: keyword, keyword KPIs, and cost, with all monetary values in RMB under the cost-per-click (CPC) model \cite{asdemir2012pricing}.
\subsection{Experiment Setup}
\textcolor{black}{Running a A/B test for keyword pruning is costly and need internal permission from advertiser. To address these challenges, we adopt a simulation-based evaluation framework that uses historical data.}

For each advertising campaign, the initial keyword set is established on day 7. KP-Agent employs a 7-day rolling window mechanism: on each day \( t \) (from \( t = 7 \) to \( t = 20 \)), it evaluates keyword performance data from days \( t - 6 \) to \( t \) to make refinement decisions for day \( t + 1 \). The daily profit acts as the market feedback for policy iteration, with the cumulative profit over the 14-day decision period (days 8--21) serving as the key performance metric. To simplify budget modeling, we assume each campaign has a fixed daily budget $B_i$,which is evenly distributed among all active keywords in the set $W_i$. When keywords are pruned, the remaining keywords receive a proportionally increased budget share, maintaining the total campaign spend. This setting allows us to simulate the effect of pruning under controlled, comparable budget conditions. The base LLM for all agents are GPT-4.1 nano.

To assess the effectiveness of KP-Agent on real-world data, we compare it with four established keyword refinement strategies:: (i) \textbf{Impression-Rank}:  This strategy ranks keywords by their 7-day average daily impression count and removes those with the lowest rankings, focusing solely on exposure potential. (ii) \textbf{CTR-Rank}: Keywords are prioritized based on their 7-day moving average click-through-rate (CTR), calculated as impressions divided by clicks. Keywords with poorest engagement are eliminated. (iii) \textbf{CVR-Rank}: This method ranks keywords by their 7-day average conversion rate (CVR), defined as conversions divided by clicks. Keywords with the poorest conversion performance are removed. (iv) \textbf{Impression Regression}: It calculates the linear regression slopes of impression values over the 7-day period. Keywords with the most negative slopes, indicating declining reach trends, are discarded.

\subsection{Main Result}
As shown in Fig.\ref{exp1}, KP-Agent achieves cumulative profit improvements ranging from 2.46\% (against the impression-based ranking baseline) to 49.28\% (against the regression-driven trend analysis method). These results highlight KP-Agent's adaptability to diverse market conditions. Notably, as the minimum keyword retention threshold $N_\text{min}$ decreases from 9 to 5, the performance gap between KP-Agent and the baseline methods widens significantly. This trend underscores KP-Agent's precision in identifying and eliminating redundant keywords. The experimental outcomes indicate that KP-Agent can effectively adaptively refine keyword sets to maximize profit, particularly when the minimum keyword requirement is lower, allowing for more aggressive removal of low-value keywords and allocating higher bid price to the remainings.

\section{Conclusion}
This paper addresses the critical yet overlooked practice of keyword pruning in sponsored search advertising (SSA) by proposing KP-Agent. Through analysis of over 0.5 million SSA records, we identified significant inefficiencies in current strategies, including resource misallocation and lack of responsiveness to market changes. KP-Agent addresses these challenges by incorporating SSA domain-specialzed toolset and a long-term memory module for few-shot learning and self-reflection. Our experiments on real-world data from Meituan demonstrate that KP-Agent outperforms baseline methods in profit generation across varying conditions, with notable improvements as the minimum keyword retention threshold decreases.

\begin{acks}
This research was supported in part by the Department of Statistics and New Asia College at The Chinese University of Hong Kong and by the DS Digital Technology Group. The research was carried out during Hou-Wan Long’s internship at Center for Digital Transformation, Cheung Kong Graduate School of Business.
\end{acks}

\section*{GenAI Usage Disclosure}
We acknowledge the use of Generative AI (GenAI) tools in the preparation of this paper as follows:
\begin{itemize}
    \item \textbf{Writing assistance:} ChatGPT (OpenAI) was used for improving grammar, rephrasing, and refining the clarity of certain paragraphs. All substantive content and structure were authored by the authors.
    \item \textbf{Code generation:} No GenAI tools were used to generate or write code used in this study.
    \item \textbf{Data processing or analysis:} No GenAI tools were used for data processing, analysis, or result generation.
\end{itemize}

All uses of GenAI tools complied with the ACM Authorship Policy on Generative AI usage.

\balance
\bibliographystyle{unsrt}
\bibliography{sample-base}

\end{document}